# INCLINED WALL PLUMES IN POROUS MEDIA

Jian-Jun SHU and Ioan POP

*School of Mechanical & Aerospace Engineering, Nanyang Technological University,*

*50 Nanyang Avenue, Singapore 639798*

## ABSTRACT
A numerical solution is presented for the natural convection from inclined wall plumes which arise from a line thermal source imbedded at the leading edge of an adiabatic plate with arbitrary tilt angle and embedded in a fluid-saturated porous medium. An appropriate transformation of the governing boundary-layer equations is proposed and a very efficient novel numerical solution is proposed to obtain rigorous numerical solutions of the transformed nonsimilar equations over a wide range of tilt angle from the vertical to the horizontal.

## INTRODUCTION

Considerable research efforts have been devoted to the study of heat transfer induced by buoyancy effects in a porous medium saturated with fluids. Interest in this convective heat transfer phenomena has been motivated by such diverse applications of the subject in contemporary technologies as thermal insulation of buildings, nuclear engineering systems, geothermal engineering, energy storage and recovery systems, storage of grain, fruits and vegetables, petroleum reservoirs, pollutant dispersion in aquifers, and catalytic reactors. There is no doubt that this area of convective heat transfer keeps attracting engineers and scientists from diversified disciplines such as mechanical engineering, chemical engineering, civil engineering, nuclear engineering, bio-engineering, food science and geothermal physics.

The steady buoyancy-induced flow arising from thermal energy sources is commonly referred to as a natural or mixed convection plume. Among such plumes, two general types may be identified - the free plume and the wall plume. The free plume is typified by the buoyant flow resulting from a point or a line source of heat. A typical wall plume is the flow resulting from a line source of heat imbedded at the leading edge of an adiabatic surface. Free and wall plumes arise in power plant steam lines, buried electrical cables, oil and gas distribution lines, volcanic eruption, disposal of nuclear wastes, hot-wire anemometry, industrial and agricultural water distribution lines, *etc*.

Natural convection resulting from a line or point source in an infinite Darcian porous medium (free plume) was first considered by Wooding (1963) and the analysis has been very much refined and generalized since then. Bejan (1978) obtained similarity solutions for the problem of axisymmetric plume from a point heat source in a porous medium on the basis of the boundary-layer approximation. The point heat sources at low Rayleigh numbers in an unbounded Darcian porous medium were investigated by Hickox (1981). The higher-order boundary layers for natural convection from a horizontal line source of heat (free plume) in a Darcian porous medium were determined by Shaw and Dawe (1985) using the method of matched asymptotic expansions. The Darcian mixed convection from a line thermal source imbedded at the leading edge of an adiabatic vertical surface (wall plume) in a saturated porous medium was numerically investigated by Kumari *et al.* (1988). Afzal and Salam (1990) considered the case when the point heat source is bounded by an adiabatic conical surface.

Coupled heat and mass transfer by natural convection at low Rayleigh number in an infinite Darcian porous medium has been considered by Ganapathy (1994) for a point source, by Larson and Poulikakos (1986) for a line source and by Lai and Kulacki (1990) for a sphere. For a large Rayleigh number, Lai (1990) obtained a similarity solution for a line source, and the closed-form solutions were presented for the special case of Lewis number equal to $1$.

Natural or mixed convection flow from free or wall plumes in non-Darcian porous media were investigated by Degan and Vasseur (1995). It was shown that the non-Darcian flow produces a much more peaked temperature profile than that predicted by the Darcian flow. Nakayama (1994) studied natural convection from a point heat source and from a line heat source in a porous medium saturated with a power-law fluid.

This paper is to investigate the Darcian boundary-layer natural convection from inclined wall plumes which arise from a line thermal source imbedded at the leading edge of an adiabatic plate with arbitrary inclination from the vertical to the horizontal and immersed in a fluid-saturated porous medium. To facilitate the analysis, a dimensionless stretched streamwise coordinate $\xi$ has been proposed. In addition, dimensionless stream function and dimensionless temperature with proper scales based on $\xi$ are defined to obtain a set of non-similar equations. Also, a new dependent constraint variable $\lambda$ has been introduced to avoid numerical integration. The variable $\xi$ also serves as an index of the inclination of the plate. For the limiting case of a vertical wall plume, $\xi = 0$, while for the limiting case of a horizontal wall plume, $\xi = 1$, respectively. In these limiting cases, the nonsimilar equations are readily reducible to the self-similar equations of the vertical and horizontal wall plumes. The new set of equations derived for arbitrary plate inclination from the vertical to the horizontal subject to an integral constraint equation of flux-conservation condition was solved by a very efficient numerical method. A considerable effort has been directed in this paper to develop a new method to solve the governing nonsimilar equations, which can be also applied to the analyses of other configurations (Shu 2004; Shu and Pop 1998a,b, 1999; Shu and Wilks 1995a,b, 1996, 2008, 2009).

## GOVERNING EQUATIONS

Here natural convection is considered from a line heat source of strength $Q$ imbedded at the leading edge of an adiabatic flat plate which is embedded in a fluid-saturated porous medium of ambient temperature $T_\infty$. The flat plate is inclined with an arbitrary angle $\psi$ to the vertical from $0$ to $\frac{\pi}{2}$, including the vertical and horizontal orientations. The physical model and coordinate system are shown in Fig. 1. Cartesian coordinates $(x, y)$ with associated velocity components $(u, v)$ are used in the subsequent analysis. The thermophysical properties of the fluid are assumed constant except for the density in the buoyancy term in the momentum equation.

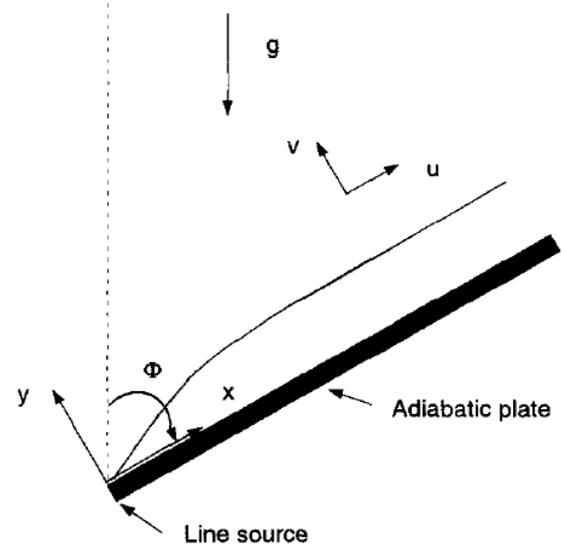

Figure 1: THE PHYSICAL MODEL AND COORDINATE SYSTEM

The governing equations for the steady, two-dimensional natural convection with the boundary layer, Boussinesq, Darcy flow and negligible inertia approximations are as follows:

$$\frac{\partial u}{\partial x} + \frac{\partial v}{\partial y} = 0,$$

$$\frac{\partial u}{\partial x} = \frac{gK\beta}{\nu}\left(\frac{\partial T}{\partial y}\cos\phi - \frac{\partial T}{\partial x}\sin\phi\right), \quad (1)$$

$$u\frac{\partial T}{\partial x} + v\frac{\partial T}{\partial y} = \alpha\frac{\partial^2 T}{\partial y^2},$$

subject to the boundary conditions

$$u = 0, \quad \frac{\partial T}{\partial y} = 0 \quad \text{at} \quad y = 0,$$

$$u \to 0, \quad T \to T_\infty \quad \text{as} \quad y \to \infty, \quad (2)$$

where $g$ is the acceleration due to gravity, $T$ is the fluid temperature, $K$ is the permeability, and $\alpha$, $\beta$ and $\nu$ are the effective thermal diffusivity, coefficient of thermal expansion and kinematic viscosity, respectively. In addition to these boundary conditions, the governing equations (1) are also subject to a constraint of energy conservation condition. That is, the total energy convected by the boundary-layer flow through the perpendicular plane at any $x > 0$ must be equal to the energy $Q$ released from the line thermal source

$$\rho C_p L \int_0^\infty u(T - T_\infty)dy = Q. \quad (3)$$

Here $\rho$, $L$ and $C_p$ denote density, length of line thermal source and specific heat of fluid.

A significant step in formulating the problem for comprehensive solutions is the introduction of the characteristic dimensionless coordinate

$$\xi = \frac{(R_a \sin\phi)^{\frac{1}{4}}}{(R_a \sin\phi)^{\frac{1}{4}} + (R_a \cos\phi)^{\frac{1}{3}}}, \quad (4)$$

where the local Rayleigh number $R_a$, is defined as

$$R_a = \frac{gK\beta T^* x}{\alpha \nu}$$

with the equivalent temperature $T^*$ given by

$$T^* = \frac{Q}{\rho C_p \alpha L}.$$

This coordinate provides the basis for a unified framework within which the features of the relative strength of the longitudinal to the transverse components of buoyant force acting on the boundary layer flow adjacent to the inclined plate.

In addition, a pseudo-similarity variable is defined as

$$\eta = \frac{y}{x}\left[(R_a \sin\phi)^{\frac{1}{4}} + (R_a \cos\phi)^{\frac{1}{3}}\right].$$

Furthermore, a dimensionless stream function and a dimensionless temperature are defined, respectively, as follows:

$$f(\xi,\eta) = \frac{y}{x}\frac{\psi(x,y)}{\alpha\eta}, \quad (5)$$

$$\theta(\xi,\eta) = \frac{x}{y}\frac{T(x,y) - T_\infty}{T^*}\eta \quad (6)$$

and

$$\lambda(\xi,\eta) = \int_0^\eta \theta \frac{\partial f}{\partial \eta} d\eta. \quad (7)$$

The differential equations and all boundary conditions transform to the following systems of equations:

$$\frac{\partial^2 f}{\partial \eta^2} = a_0(\xi)\frac{\partial \theta}{\partial \eta} + b_0(\xi)\frac{\partial \theta}{\partial \xi} + c_0(\xi)\eta\frac{\partial \theta}{\partial \eta} + d_0(\xi)\theta,$$

$$\frac{\partial^2 \theta}{\partial \eta^2} + p_0(\xi)\frac{\partial (f\theta)}{\partial \eta} = q_0(\xi)\left(\frac{\partial \theta}{\partial \eta}\frac{\partial f}{\partial \xi} - \frac{\partial f}{\partial \eta}\frac{\partial \theta}{\partial \xi}\right) \quad (8)$$

$$\frac{\partial \lambda}{\partial \eta} = \theta \frac{\partial f}{\partial \eta},$$

with boundary conditions

$$f|_{\eta=0} = \frac{\partial \theta}{\partial \eta}\Big|_{\eta=0} = \lambda|_{\eta=0} = 0, \quad \frac{\partial f}{\partial \eta}\Big|_{\eta \to \infty} = 0, \quad \lambda|_{\eta \to \infty} = 1. \quad (9)$$

Here

$$a_0(\xi) = (1-\xi)^3, \quad b_0(\xi) = \frac{\xi^5(1-\xi)}{12}, \quad c_0(\xi) = \frac{\xi^4(8+\xi)}{12}$$

$$d_0(\xi) = \frac{\xi^4(4-\xi)}{12}, \quad p_0(\xi) = \frac{4-\xi}{12}, \quad q_0(\xi) = \frac{\xi(1-\xi)}{12}.$$

The constraint condition $\lambda|_{\eta \to \infty} \to 1$ implies $\theta \frac{\partial f}{\partial \eta}\Big|_{\eta \to \infty} \to 0$, that is, either $\theta|_{\eta \to \infty} \to 0$ or $\frac{\partial f}{\partial \eta}\Big|_{\eta \to \infty} \to 0$. The boundary condition $\theta|_{\eta \to \infty} \to 0$ is dropped to avoid computational overdetermination.

For the limiting case of a vertical wall plume, $\xi = 0$, equations (8) are readily reduced to the following self-similar equations:

$$\frac{\partial^2 f}{\partial \eta^2} = \frac{\partial \theta}{\partial \eta},$$

$$\frac{\partial^2 \theta}{\partial \eta^2} + \frac{1}{3}\frac{\partial (f\theta)}{\partial \eta} = 0, \quad (10)$$

$$\frac{\partial \lambda}{\partial \eta} = \theta \frac{\partial f}{\partial \eta}.$$

The analytical solution of these equations is given by

$$f(\eta) = 3^{\frac{2}{3}} \tanh\left(\frac{\eta}{23^{\frac{1}{3}}}\right), \quad \theta(\eta) = \frac{3^{\frac{1}{3}}}{2\cosh^2\left(\frac{\eta}{23^{\frac{1}{3}}}\right)},$$

$$\lambda(\eta) = \frac{1}{2}\tanh\left(\frac{\eta}{23^{\frac{1}{3}}}\right)\left[3 - \tanh^2\left(\frac{\eta}{23^{\frac{1}{3}}}\right)\right]. \quad (11)$$

For the other limiting case of a horizontal wall plume, $\xi = 1$, equations (8) are readily reduced to the following self-similar equations:

$$\frac{\partial^2 f}{\partial \eta^2} = \frac{3}{4}\eta\frac{\partial \theta}{\partial \eta} + \frac{1}{4}\theta,$$

$$\frac{\partial^2 \theta}{\partial \eta^2} + \frac{1}{4}\frac{\partial (f\theta)}{\partial \eta} = 0, \quad (12)$$

$$\frac{\partial \lambda}{\partial \eta} = \theta \frac{\partial f}{\partial \eta}.$$

## NUMERICAL METHOD

The solution to the nonsimilar equations (8) of the inclined wall plume in a porous medium with boundary conditions (9) is obtained numerically using finite differences.

Now, the equations are written as a first order system by introducing the new dependent variables $q(\xi,\eta)$ and $w(\xi,\eta)$ as follows:

$$\frac{\partial f}{\partial \eta} = q,$$

$$\frac{\partial q}{\partial \eta} = a_0(\xi)w + b_0(\xi)\frac{\partial \theta}{\partial \xi} + c_0(\xi)\eta w + d_0(\xi)\theta,$$

$$\frac{\partial \theta}{\partial \eta} = w, \quad (13)$$

$$\frac{\partial w}{\partial \eta} = -p_0(\xi)(fw + \theta q) + q_0(\xi)\left(w\frac{\partial f}{\partial \xi} - q\frac{\partial \theta}{\partial \xi}\right),$$

$$\frac{\partial \lambda}{\partial \eta} = \theta q.$$

The boundary conditions now become

$$f|_{\eta=0} = w|_{\eta=0} = \lambda|_{\eta=0} = 0, \quad q|_{\eta\to\infty} = 0, \quad \lambda|_{\eta\to\infty} = 1.$$

An arbitrary rectangular net of points $(\xi_n, \eta_j)$ on $0 \leq \xi \leq 1$, $\eta \geq 0$ is placed and the notation is used as:

$$\xi_0 = 0, \quad \xi_n = \xi_{n-1} + k_n, \quad n = 1,2,\cdots,K, \quad \xi_K = 1;$$
$$\eta_0 = 0, \quad \eta_j = \eta_{j-1} + h_j, \quad j = 1,2,\cdots,J; \quad (14)$$

If $(f_j^n, q_j^n, \theta_j^n, w_j^n)$ are to approximate $(f, q, \theta, w)$ at $(\xi_n, \eta_j)$, the difference approximations are defined, for $1 \leq j \leq J$, by

$$\frac{f_j^n - f_{j-1}^n}{h_j} = q_{j-1/2}^n,$$

$$\frac{q_j^{n-1/2} - q_{j-1}^{n-1/2}}{h_j} = a_{n-1/2} w_{j-1/2}^{n-1/2} + b_{n-1/2} \frac{\theta_{j-1/2}^n - \theta_{j-1/2}^{n-1}}{k_n}$$
$$+ c_{n-1/2} \eta_{n-1/2} w_{j-1/2}^{n-1/2} + d_{n-1/2} \theta_{j-1/2}^{n-1/2}$$

$$\frac{\theta_j^n - \theta_{j-1}^n}{h_j} = w_{j-1/2}^n, \quad (15)$$

$$\frac{w_j^{n-1/2} - w_{j-1}^{n-1/2}}{h_j} = -p_{n-1/2}(fw + \theta q)_{j-1/2}^{n-1/2} + q_{n-1/2}$$

$$\times \left( w_{j-1/2}^{n-1/2} \frac{f_{j-1/2}^n - f_{j-1/2}^{n-1}}{k_n} - q_{j-1/2}^{n-1/2} \frac{\theta_{j-1/2}^n - \theta_{j-1/2}^{n-1}}{k_n} \right)$$

$$\frac{\lambda_j^n - \lambda_{j-1}^n}{h_j} = (\theta q)_{j-1/2}^n,$$

where $\xi_{n-1/2} = (\xi_n + \xi_{n-1})/2$, $a_{n-1/2}$, $b_{n-1/2}$, $c_{n-1/2}$, $d_{n-1/2}$, $p_{n-1/2}$ and $q_{n-1/2}$ are the values of $a_0(\xi)$, $b_0(\xi)$, $c_0(\xi)$, $d_0(\xi)$, $p_0(\xi)$ and $q_0(\xi)$ at $\xi_{n-1/2}$, respectively, and for any function $z(\xi, \eta)$, a notation is introduced for averages and intermediate values as

$$z_{j-1/2}^n = (z_j^n + z_{j-1}^n)/2, \quad z_j^{n-1/2} = (z_j^n + z_j^{n-1})/2,$$
$$z_{j-1/2}^{n-1/2} = (z_j^n + z_{j-1}^n + z_j^{n-1} + z_{j-1}^{n-1})/4.$$

Note that the first, third and last equations (15) are centered at $(\xi_n, \eta_{j-1/2})$, while the second and fourth equations (15) are centered at $(\xi_{n-1/2}, \eta_{j-1/2})$, i.e. when a $\xi$ derivative is absent, equations can be differenced about the point $(\xi_n, \eta_{j-1/2})$. It is found in practice that this damps high-frequency Fourier error components better than differencing about $(\xi_{n-1/2}, \eta_{j-1/2})$.

The boundary conditions become simply:

$$f_0^n = 0, \quad w_0^n = 0, \quad \lambda_0^n = 0, \quad q_J^n = 0, \quad \lambda_J^n = 1. \quad (16)$$

The nonlinear difference equations may now be solved recursively starting with $n = 0$ (on $\xi = \xi_0 = 0$). In the case of $n = 0$, the first, third and last equations (15) and (16) are retained with $n = 0$ and the second and fourth equations (15) are simply altered by setting $\xi_{n-1/2} = 0$ and using superscripts $n = 0$ rather than $n = 1/2$ in the remaining terms. The resultant difference equations are then solved by Newton iteration. Solutions are obtained on different sized grids and Richardson's extrapolation used to produce results of high accuracy.

## EXTRAPOLATING THE RESULTS

Since central difference is used, the exact numerical solution of the difference equations (15) and (16) is a second-order accurate approximation. The local truncation errors of this difference scheme can be written as a Taylor series in powers of $h^2$ and $k^2$ where $h = \max_j h_j$ and $k = \max_n k_n$. It is therefore possible, by solving the problem on different sized grids and using Richardson's extrapolation, to produce results of high accuracy provided the truncation errors are larger than the iteration errors. For example, each cell of the net (14) is divided into $m$ subintervals both in the $\xi$ direction and in the $\eta$ direction where $m$ is an integer. The problem is solved numerically for $m = 1$, $2$, $3$ and $4$. If $z_m$ denotes the results of any actual variable function $z(\xi, \eta)$ at a common grid point then $z_m$ has accuracy $O(h^2 + k^2)$. Since the truncation error is proportional to the square of $h$ and $k$ then

$$z_{12} = (4z_2 - z_1)/3, \quad z_{23} = (9z_3 - 4z_2)/5,$$
$$z_{34} = (16z_4 - 9z_3)/7$$

have errors $O(h^4 + k^4)$ and

$$z_{123} = (9z_{23} - z_{12})/8, \quad z_{234} = (16z_{34} - 4z_{23})/12$$

will be in error by $O(h^6 + k^6)$ and, finally,

$$z_{1234} = (16z_{234} - z_{123})/15.$$

The results quoted are $z_{1234}$ and error is estimated by maximum of the difference $|z_{1234} - z_{234}|$, which being a global error estimate measures the actual error in $z$.

In order to assess accuracy, the numerical results for a vertical wall ($\phi = 0$, i.e. $\xi = 0$) are compared with the analytical results given by equation (11). Thus, the value of $\theta(0)$ both from equations (8) and (9) and from equation (11) when $\xi = 0$ is $\theta(0) = 0.72112478$, which shows that the agreement is excellent. It is, therefore, strongly confident that the present numerical scheme works very efficient.

## RESULTS AND DISCUSSION

The application of the results obtained in this paper lies in the determination of the velocity and temperature profiles in the boundary layer as well as the wall velocity and wall temperature profiles. The dimensionless velocity $\frac{u}{\alpha/x}$ and temperature $\frac{T - T_\infty}{T^*}$ profiles over the dimensionless transverse coordinate $\frac{y}{x}$ are depicted in Figs. 2 and 3 for different values of the tilt angle $\phi$ and $R_a = 10^3$. It can see that the maximum of the velocity and temperature profiles occurs at the wall $y = 0$. In addition, the longitudinal velocity profiles decrease near the wall as $\phi$ increases, while the temperature profiles increase monotonically with increasing values of $\phi$; the fluid adjacent to the line source becomes hotter, and it begins to rise until it reaches a maximum. This is due to the decrease of the component of buoyancy force with increasing the tilt angle $\phi$. Thus, for a fixed angle $\phi$, the temperature increases near the wall and decreases more rapidly towards the outer edge of the plume and the width of the plume becomes less extensive. Therefore, the boundary and convective effects are more important for a wall plume than for a free plume.

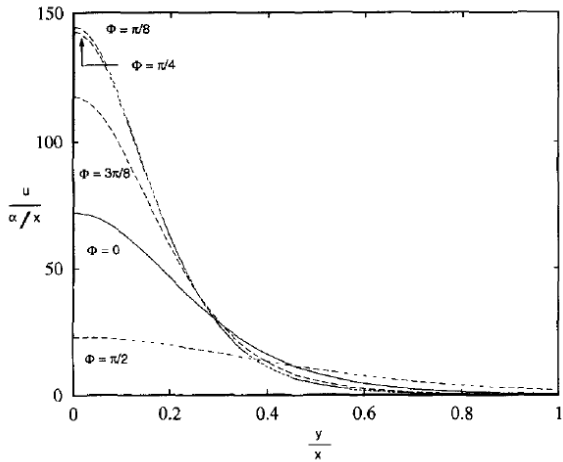

Figure 2: REPRESENTATIVE VELOCITY PROFILES FOR $R_a = 10^3$

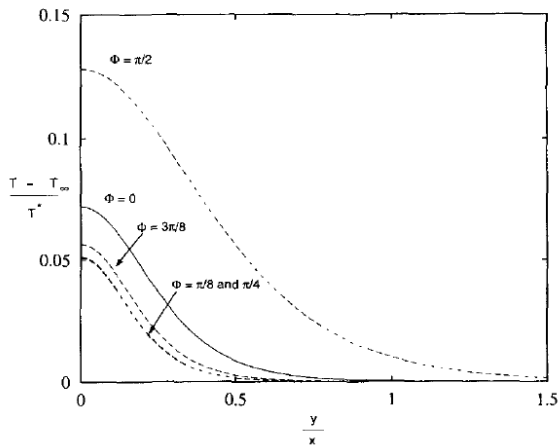

Figure 3: REPRESENTATIVE TEMPERATURE PROFILES FOR $R_a = 10^3$

The maximum dimensionless velocity and temperature profiles (*i.e.* wall velocity and wall temperature) are shown in Figs. 4 and 5 versus the tilt angle $\phi$, *i.e.* over the range of $0 \leq \xi \leq 1$ and some values of $R_a$. As expected, the wall velocity increases, while the wall temperature decreases with increasing the parameter $\phi$. This is because a more vigorous flow when the plate is tilted away from the horizontal. It is also noticed from Fig. 4 that the wall velocity increases, with the increase of the Rayleigh number $R_a$. This causes a reduction of the wall temperature when $R_a$ is increased (see, Fig. 5). Therefore, at larger values of $R_a$, the plume becomes thicker, resulting in increased entrainment of fluid in the boundary layer from the outer region. The entrainments depends on the extent of outer region, *i.e.* whether the fluid can entrain from entire or limited space. Furthermore, Figs. 4 and 5 show that the longitudinal velocity and temperature profiles at the wall do not change monotonically with the tilt angle $\phi$. Highest wall velocity and lowest wall temperature can be found at $\phi \approx \dfrac{\pi}{6}$. This may be due to the combined effects between transverse buoyancy and longitudinal buoyancy. This observation could be useful for engineering applications.

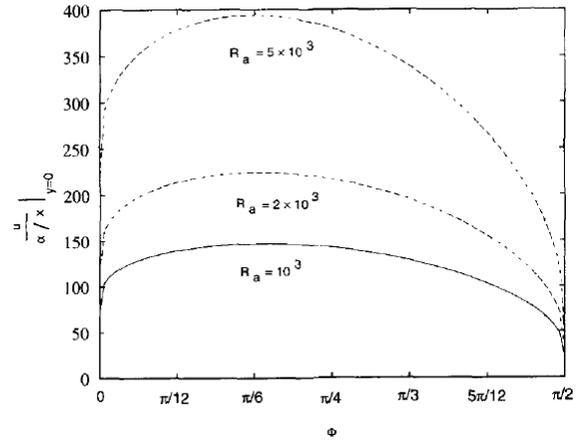

Figure 4: VARIATIONS OF WALL VELOCITY WITH $\phi$

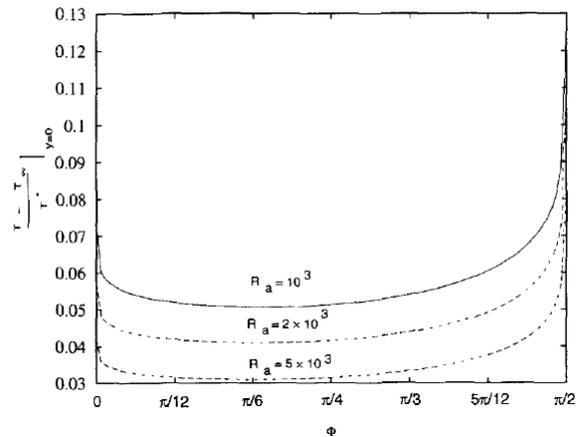

Figure 5: VARIATIONS OF WALL TEMPERATURE WITH $\phi$

It is concluded by noting that this flow configuration is of considerable importance in technology and the present work considers the resultant boundary-layer flow in order to determine the velocity and temperature profiles, as well as other physical aspects of interest. The results of the present study are important in technology, for example, in the positioning of components dissipating energy on vertical circuit boards placed in a porous medium and in the positioning of the boards themselves. Heat transfer and natural convection flow considerations are very important in this area and also in several frequently encountered manufacturing processes. The interaction of the flows arising from several elements, which constitute steady thermal sources, on a vertical insulating surface is an important problem and further work needs to be done on it in order to determine the nature of the resultant flow and heat transfer.